# Transition-metal distribution in kagome antiferromagnet $CoCu_3(OH)_6Cl_2$ revealed by resonant x-ray diffraction

Yuesheng Li[1], Jianlong Fu[2], Zhonghua Wu[2] and Qingming Zhang[1,3]

[1] Department of Physics, Renmin University of China, Beijing 100872, P. R. China.

[2] Institute of High Energy Physics Chinese Academy of Science, Beijing 100049, P. R. China.

[3] Corresponding author. Tel. /Fax.: +86-10-62517767. E-mail address: qmzhang@ruc.edu.cn

**Abstract.** The distribution of chemically similar transition-metal ions is a fundamental issue in the study of herbertsmithite-type kagome antiferromagnets. Using synchrotron radiation, we have performed resonant powder x-ray diffractions on newly synthesized $CoCu_3(OH)_6Cl_2$, which provide an exact distribution of transition-metal ions in the frustrated antiferromagnet. Both magnetic susceptibility and specific heat measurements are quantitatively consistent with the occupation fractions determined by resonant x-ray diffraction. The distribution of transition-metal ions and residual magnetic entropy suggest a novel low temperature (T < 4 K) magnetism, where the interlayer triangular spins undergo a spin-glass freezing while the kagome spins still keep highly frustrated.

## 1. Introduction

Spin frustration is one of the most fascinating but challenging issues in condensed matter and many novel spin ground states have been revealed in spin-frustrated systems. Among the novel ground states, the so-called spin liquid, which shows no long-range order in spin correlations even down to 0 K, has attracted particular interests because it was proposed to be ultimately connected with high-temperature superconductivity in cuprates [1-5]. Spin liquid is expected only in a highly spin-frustrated system, such as kagome lattice. However, the real compounds with perfect kagome planes are quite rare. Recently herbertsmithite-type $S = 1/2$ kagome antiferromagnet $ZnCu_3(OH)_6Cl_2$ was discovered [6,7]. Due to its low dimensionality, low spin quantum number and highly geometrically frustrated configuration (kagome lattice of corner-sharing equilateral triangles), it was considered as the best candidate in exploring spin liquid and other spin states [8]. Thermodynamics measurements indicate that no long-range magnetic order is established down to 50 mK [9], exactly as required by kagome antiferromagnetic spin-1/2 Heisenberg model (KAHM) [1,10-13].

Ideally $Zn^{2+}$ and $Cu^{2+}$ ions are expected to occupy interlayer -3M(100) sites (triangular sites, coordination number 6) and in-plane 2/M(100) sites (kagome sites, coordination number 4), respectively. Unfortunately, site-exchange between $Zn^{2+}$ and $Cu^{2+}$ actually may exist in real materials as the two ions have very similar ionic radii and the energy barrier built by Jahn-Teller distortion is too small to completely prevent the site exchange. The Zn-Cu site mixture in $ZnCu_3(OH)_6Cl_2$ was confirmed by many experiments [14,15]. It has a significant influence on magnetism in the kagome system. For example, it may partially release spin frustration and induce a third-dimensional (interlayer) exchange coupling. Therefore, quantitative determination of site exchange becomes a crucial issue for exploring the kagome spin system.

Recently we successfully synthesized single-phased herbertsmithite-type $CoCu_3(OH)_6Cl_2$ by hydrothermal method with rotating pressure vessels [16]. Dramatically different from $ZnCu_3(OH)_6Cl_2$, $CoCu_3(OH)_6Cl_2$ exhibits a spin-glass transition at low temperatures ( < 4 K ). At the same time, $ZnCu_3(OH)_6Cl_2$-like crystal symmetry (R-3m) is proved to be preserved and no magnetic hysteresis is observed down to 2 K in $CoCu_3(OH)_6Cl_2$. If the amount of interlayer $Cu^{2+}$ ions is higher than a critical level, magnetic hysteresis at low temperatures will be seen due to the formation of locally distorted $Cu_2(OH)_3Cl$ ferromagnetic phases [17-19]. Actually it has been observed in $Zn_xCu_{4-x}(OH)_6Cl_2$ and $Mg_xCu_{4-x}(OH)_6Cl_2$ (x < 1) [20-22]. Contrastly this suggests that only a small amount of $Cu^{2+}$ ions enter into interlayer sites in $CoCu_3(OH)_6Cl_2$. But the qualitative estimation is not enough to understand novel magnetism in kagome antiferromagnet. A further quantitative knowledge of the distribution of transition-metal ions is highly required, just as in the case of $ZnCu_3(OH)_6Cl_2$ [14].

X-ray diffraction may be the most direct and effective method for the purpose. However a conventional x-ray diffraction with Cu target fails to do this, as the normal x-ray atomic scattering factor $f_0(Q = 0)$ of Co (26.98900) is very close to that of Cu (28.98590) under a non-resonance condition, [23] due to their small difference in atomic number. On the other hand, the scattering factors of Co and Cu are significantly different in a resonance case. In this case, the factor can be described with a complex form $f = f_0(Q) + f'(E)+i*f''(E)$, where $f_0(Q)$ is the non-resonant atomic scattering factor, $f'(E)$ and $f''(E)$ are the energy-dependent anomalous (real) and absorption (imaginary) parts of delta-form factor, respectively. For instance, at Co K-edge (7711 eV), the form factors are $f_{Co}'$ (7711 eV) ~ -7.790; $f_{Co}''$ (7711 eV) ~ 0.475 and $f_{Cu}'$ (7711 eV ) ~ -1.686; $f_{Cu}''$ (7711 eV) ~ 0.639, while at Cu K-edge (8978 eV), $f_{Cu}'$ (8978 eV) ~ -7.854; $f_{Cu}''$ (8978 eV) ~ 0.483 and $f_{Co}'$ (8978 eV) ~ -0.849; $f_{Co}''$ (8978 eV) ~ 3.032 [14, 24]. Therefore, the difference in atomic scattering factor can be effectively enlarged under proper resonance conditions, which allows one to clearly distinguish Co ions from Cu ions by performing resonant x-ray diffraction [25-29]. Then their reliable occupation fractions at different sites can be given by combined Rietveld refinements based on resonant x-ray diffraction patterns.

In this paper, we have performed x-ray diffraction measurements with synchrotron radiation at three energies of Co K-edge 7711 eV, Cu K-edge 8978 eV and 12.7 keV. We found a significant difference in relative intensities for three diffraction patterns. For all the three patterns, careful combined Rietveld refinements are carried out. That means all the refinements share the same refining parameters, such as lattice parameters, thermal motion parameters, atomic coordinates, atomic site fractions and profile function. Completely consistent results are produced by the refinements, which demonstrate that the occupation fractions are ~ 93.6% and 80.0% for Cu at kagome sites and Co at interlayer site. The distribution of transition-metal ions determined by resonant x-ray diffraction is quantitatively confirmed by our magnetization and specific heat measurements. More important, the measurements suggest the spin-glass transition at low temperature (T ~ 3.5 K), which may be caused by a freezing of interlayer triangular spins while kagome spins still keep highly frustrated, maybe in spin liquid states as herbertsmithite. As a result, resolving the occupation fractions of transition-metal ions provides a solid starting point to explore magnetism in the kagome spin system.

2. Materials and methods

$CoCu_3(OH)_6Cl_2$ powder sample was synthesized under hydrothermal conditions in a homogeneous reactor with continuously rotating pressure vessel. The detailed procedure on synthesis was described elsewhere [16]. In the end product, no sign of secondary phases such as $Co(OH)_2$, CoO, and CuO, was determined by both synchrotron and laboratory (Shimadzu XRD-7000) x-ray diffraction. Magnetic and specific heat measurements were made with a Physical Property Measurement System (PPMS) by Quantum Design.

Synchrotron x-ray diffraction was performed in the diffraction experimental station (4B9A) of Beijing Synchrotron Radiation Facility (BSRF). The powder sample was impacted into the 1cm×1cm×1.5mm square tank of a slide glass. The highly monochromatic x-ray from Si (111) double-crystal monochromator was focused on the surface of the powder sample with a 3×1 mm$^2$ light spot. Three diffraction data, at Co K-edge (7711 eV, 1.608 Å), 12.7 keV (0.979 Å) and Cu K-edge (8978 eV, 1.381 Å) modes, were collected at a successive process without changing external conditions like sample temperature (295 K) and sample position. All of the diffraction data was processed and fitted using Rietveld techniques with the General Structure Analysis System (GSAS) program. GSAS technical manual indicates that for X-ray data the correction by absorption coefficient is suitable only for data taken in the Debye-Scherrer geometry and it should not be used for Bragg-Brentano or other flat plate geometries [23]. We obtained the constant wavelength (CW) X-ray data in a powder diffractometer with Bragg-Brentano geometry, so the self-absorption correction was not taken into account in the refinement process.

## 3. Results and discussion

Diffraction patterns at three incident photon energies are shown in Fig. 1. For most Bragg reflections, one can find obvious intensity modulations with incident energies. The Bragg peaks with larger modulations are highlighted with dashed boxes in Fig. 1. Actually intensity modulations are clear responses to occupation preference of transition-metal ions. In the following we will have a qualitative discussion on this point. Later we will perform combined Rietveld refinements to quantitatively study the Co/Cu occupation at kagome and triangular sites.

In order to analyze the implications of intensity variations, we need to know structure factors of different sites. For interlayer triangular site -3M(100), there are three symmetrical positions per unit cell, i.e., (0, 0, 1/2), (1/3, 2/3, 1/6) and (2/3, 1/3, 5/6). Its complex structure factor can be written as

$$F_t^{(hkl)} = \sum_{j=1}^{3} \exp[i \cdot 2\pi(h \cdot x_j + k \cdot y_j + l \cdot z_j)] \times (X_t^{Co} \cdot f_{Co} + X_t^{Cu} \cdot f_{Cu}) = \beta_t \times f_t \quad (1)$$

where $X_t^{Co}$ and $X_t^{Cu}$ are site occupation fractions of Co and Cu at triangular sites respectively and $X_t^{Co} + X_t^{Cu} = 1$. In-plane kagome site 2/M(100) has nine symmetrical positions, which are (1/2, 0, 0), (0, 1/2, 0), (1/2, 1/2, 0), (1/6, 5/6, 1/3), (1/6, 1/3, 1/3), (2/3, 5/6, 1/3), (1/3, 1/6, 2/3), (5/6, 1/6, 2/3) and (5/6, 2/3, 2/3). Similarly, its structure factor is

$$F_k^{(hkl)} = \sum_{j=1}^{9} \exp[i \cdot 2\pi(h \cdot x_j + k \cdot y_j + l \cdot z_j)] \times (X_k^{Co} \cdot f_{Co} + X_k^{Cu} \cdot f_{Cu}) = \beta_k \times f_k \quad (2)$$

where $X_k^{Co}$ and $X_k^{Cu}$ are site occupation fractions of Co and Cu at kagome sites and $X_k^{Co} + X_k^{Cu} = 1$. Simple calculations demonstrate that the possible value for $\beta_t$ is 3 or -3 while -3 or 9 for $\beta_k$. The contributions from all the sites occupied by O and Cl are written as $\sum \beta f_R$. The structure factors are connected with relative intensities through $I \propto |\beta_t f_t + \beta_k f_k + \sum \beta f_R|^2$. The observed relative intensities of Bragg reflections at three incident energies and structure factors are summarized in Table 1, in which the contributions from O and Cl are also included [16, 23].

Suppose that $Co^{2+}$ and $Cu^{2+}$ have no occupation preference for triangular or kagome sites,

we would expect $f_t = f_k$. Then for the reflections with $(\beta_t, \beta_k) = (3, -3)$, the contributions to structure factors from the two kinds of sites would be approximately cancelled. That means the relative intensities of those reflections would keep almost unchanged for any incident energies. However, it is contrast to what we have observed in $CoCu_3(OH)_6Cl_2$. As shown in Table 1 and Fig. 1, the relative intensities of the peaks with $(\beta_t, \beta_k) = (3, -3)$ exhibit significant modulations with incident energies, like (012), (110) etc. This suggests $Co^{2+}/Cu^{2+}$ has a strong site occupation preference for triangular/kagome site. It can be understood with strong Jahn-Teller effect. Jahn-Teller-inactive $Co^{2+}$ ions favor six-coordinated interlayer triangular sites -3M(100) while Jahn-Teller-active $Cu^{2+}$ ions prefer to occupy four-coordinated kagome sites 2/M(100) [6]. In fact, not only the above Bragg reflections exhibit intensity evolution with incident energies. If carefully comparing the three patterns in Fig. 1, we find that other reflections show similar relative intensity changes with photon energies, as summarized in Table 1. Using non-resonant atomic scattering factor $f_0$ listed in Table 1 and considering resonant effects (f' ~ -8), we can make a rough estimation on relative intensities for different photon energies. We find that it well explains the observed intensity evolution tendency of each reflection. Actually we can quantitatively and consistently determine occupation fractions from intensity variations, as we will see in the following refinements. We will demonstrate that resonant x-ray diffraction is a powerful tool in resolving the distribution of Co and Cu in $CoCu_3(OH)_6Cl_2$.

To quantitatively determine site occupation, we need to make combined Rietveld refinements based on three diffraction patterns. This means that the values of refining parameters are shared in all refinements, such as lattice parameters (a and c), thermal motion parameters ($U_{iso}$), atomic coordinates, occupation fractions and parameters of the profile function etc. It is a rigorous examination of self-consistency of diffraction data and warrants the reliability of refined occupation fractions. We employ GSAS package [23] to do the refinements. The theoretical delta-form factors of Co, Cu, O and Cl are used in our refinements. The theoretical values are accurate as warranted by experiments [14].

Combined Rietveld refinement plots are shown in Fig. 2. All the reflections are well assigned in the refinements. No additional reflection is found, indicating no sign of secondary phase. Refinement parameters are summarized in the appendix. The total residuals of 3~5% indicate that the refinements are of high-quality. The refinements give an exact compound formula to be $Co_{0.992(10)}Cu_{3.008(10)}(OH)_6Cl_2$, which is very close to the nominal one ($CoCu_3(OH)_6Cl_2$) and corresponding to the ICP result, Co:Cu = 1.06(4):2.94. Co occupation fraction in kagome layers $X_k^{Co}$, is determined to be 6.4(1.0)% from the combined refinements. In other words, 93.6(1.0)% kagome sites are occupied by $Cu^{2+}$ ions ($X_k^{Cu}$). Meanwhile, Co occupation fraction at interlayer triangular sites ($X_t^{Co}$) is derived to be 80.0(1.0)% and the rest (~ 20%) are occupied by Cu. The occupation is at the same level as in herbertsmithite, where the amount of inter-site mixing defects is estimated to be 5 - 10% [15,31].

The occupation fractions are supported by our thermodynamic experiments. We have measured magnetic susceptibilities under zero-field cooling, which are shown in Fig. 3. In the whole temperature range (2 – 300 K), magnetic susceptibilities of $CoCu_3(OH)_6Cl_2$ are much larger than those of $ZnCu_3(OH)_6Cl_2$. Clearly this is caused by quasi-free $Co^{2+}$ spins at interlayer triangular sites. To approximately estimate the amount of quasi-free $Co^{2+}$ spins, we need to filter out the contributions from kagome layers and $Cu^{2+}$ spins at interlayer triangular sites. Roughly these contributions can be represented by magnetic susceptibilities of $ZnCu_3(OH)_6Cl_2$, considering the similar occupation condition in both compounds as mentioned above. The subtracted magnetic susceptibilities turn out to be a perfect Currie-like tail above 5 K, as shown in the inset of Fig. 3. The fitted Curie Weiss temperature, - 8.11(8) K, suggests a weak magnetic coupling between triangular ($Co^{2+}$) and neighboring kagome ($Cu^{2+}$) spins. This coincides with the refined bond angle Co-O-Cu of ~ 96°, which is very close to the critical bond angle of 95° between ferromagnetic and antiferromagntic exchange interactions [32,33]. In $CoCu_3(OH)_6Cl_2$, $Co^{2+}$ dominantly locates at interlayer triangular site to form $Co-(OH)_6$ octahedral (coordination number 6), as revealed by the above resonant x-ray diffraction analysis. In this kind of crystal field, $Co^{2+}$ has a high-spin configuration with $\mu_{eff}$ = 4.7 ~ 5.2 $\mu_B$ [34]. Using the momemt and fitted Curie constant C = 0.842(4) Kcm$^3$/mol Cu, we can estimate Co occupation fraction at triangular sites to be 75 ~ 91 %, which is in good agreement with the above refinement results (80.0(1.0)%).

We have further carried out specific heat measurements from 2 to 50 K. The temperature dependence of specific heat is shown in the inset of Fig. 4. We extract magnetic contributions by subtracting lattice contributions (Fig. 4). Fitting the low temperature magnetic specific heat to a power-law $\gamma T^{\alpha}$, we get $\alpha= 1.58(1)$, which implies low-dimensional magnetic correlations. The variation of magnetic entropy $\Delta S_M$ between 0 and $\geq 30$ K, is ~1.86Rln2 per mol formula. Interlayer quasi-free spin contribution to magnetic entropy is $X_t^{Co} R \ln 4 + X_t^{Cu} R \ln 2$. If taking the occupation fractions given by the above refinements, we will have $\Delta S_M$ ~ 1.8(1)Rln2. While for kagome spins, $3 \cdot (X_k^{Co} R \ln 4 + X_k^{Cu} R \ln 2)$ ~ 3.2(1)Rln2. These indicate that the spin-glass transition at low temperature (~ 3.5 K) may occur among interlayer triangular spins while the kagome spins still keep highly frustrated without any freezing as in herbertsmithite. The high consistency between resonant x-ray diffraction and magnetic specific heat, once again demonstrates that the occupation fractions are reliable.

## 4. Conclusions

We employ synchrotron resonant x-ray diffraction at different incident photon energies, Co K-edge, 12.7 keV ($\lambda= 0.979$ Å) and Cu K-edge, to identify the distribution of transition-metal ions in the newly synthesied kagome compound $CoCu_3(OH)_6Cl_2$. The diffraction patterns show significant modulations in relative intensities with different incident energies, which strongly suggests that there exists site occupation preference of $Co^{2+}$ and $Cu^{2+}$ driven by Jahn-Teller effect. Combined Rietveld refinements demonstrate that kagome sites are occupied by $Cu^{2+}$ ions at a percent of 93.6(1.0)% and $Co^{2+}$ ions occupy triangular sites at a percent of 80.0(1.0)%. The occupation fractions are further confirmed by magnetic susceptibilities and specfic heat measurements. Although having the same crystal symmetry as $ZnCu_3(OH)_6Cl_2$ and similar occupation fractions of transition-metal ions, $CoCu_3(OH)_6Cl_2$ exhibits a different low-temperature magnetism. The resolved distribution of transion-metal ions provides a good basis for further exploration on novel magnesim in the new kagome compound.

**Acknowledgements**

This work was supported by the NSF of China, the 973 program (Grant Nos. 2011CBA00112 & 2012CB921701), the Fundamental Research Funds for the Central Universities and the Research Funds of Renmin University of China.

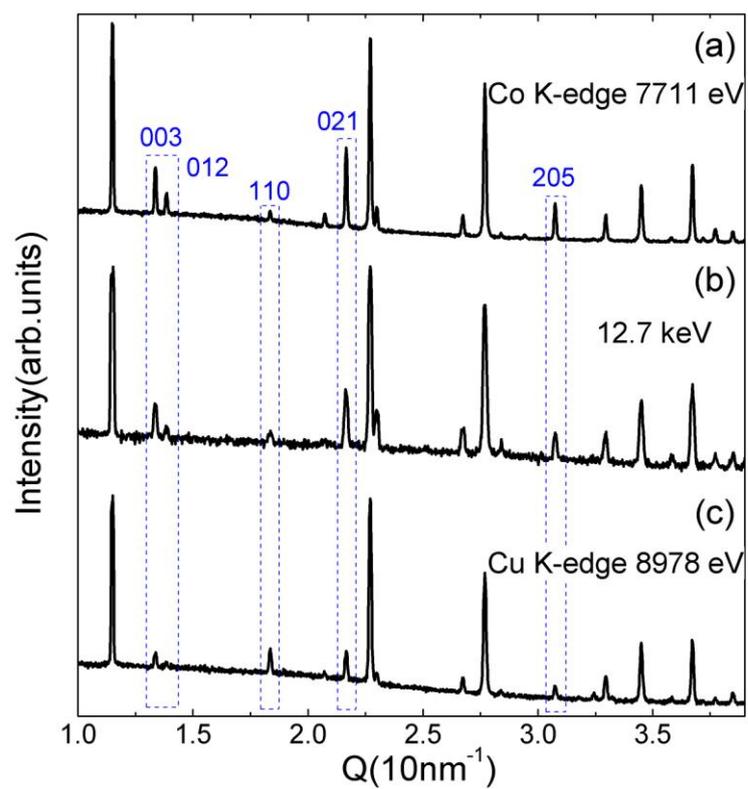

**Fig. 1.** X-ray powder diffraction patterns obtained at (a) Co K-edge (b) 12.7 keV (λ= 0.979 Å ) and (c) Cu K-edge for $CoCu_3(OH)_6Cl_2$, where Q = 2π/d = 4πsinθ/λ. The Bragg reflections with larger intensity variations are highlighted with blue dashed boxes for comparison.

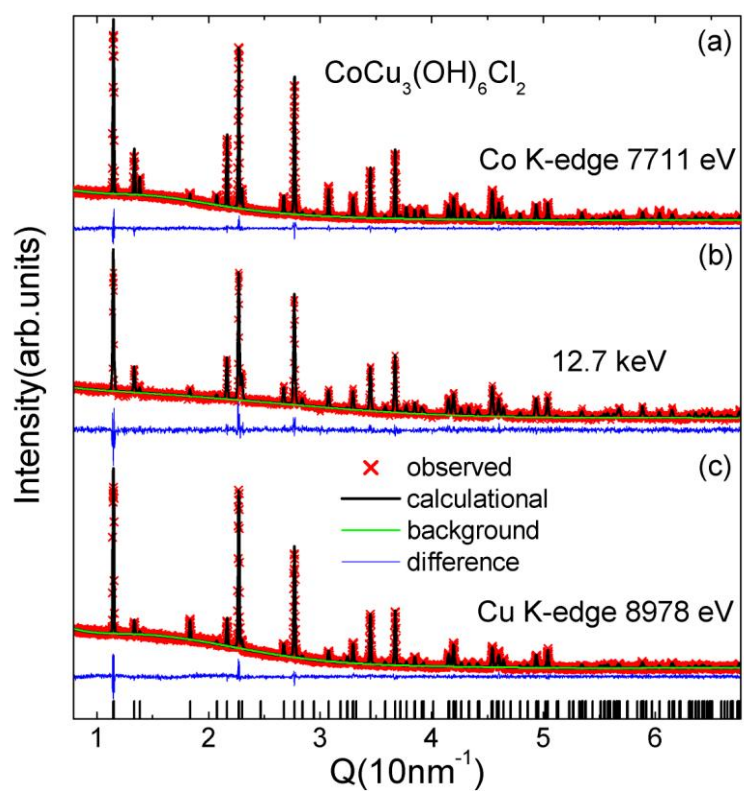

**Fig. 2.** X-ray powder diffraction patterns and combined Rietveld refinements for incident x-ray photon energy of (a) Co K-edge (b) 12.7 keV (λ= 0.979 Å) and (c) Cu K-edge.

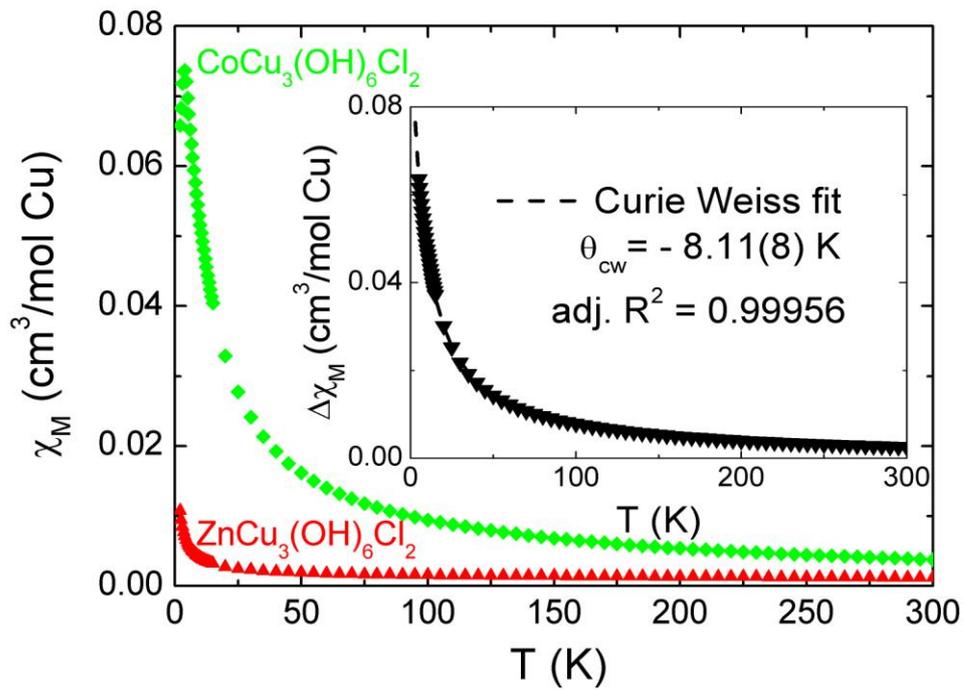

**Fig. 3.** Temperature dependence of magnetic susceptibilities in $CoCu_3(OH)_6Cl_2$ and $ZnCu_3(OH)_6Cl_2$ measured under an applied field of 2000 Oe. Inset: Subtracted susceptibilities of $CoCu_3(OH)_6Cl_2$ by those of $ZnCu_3(OH)_6Cl_2$, and the curve is a Curie-Weiss fit from 5 to 300 K.

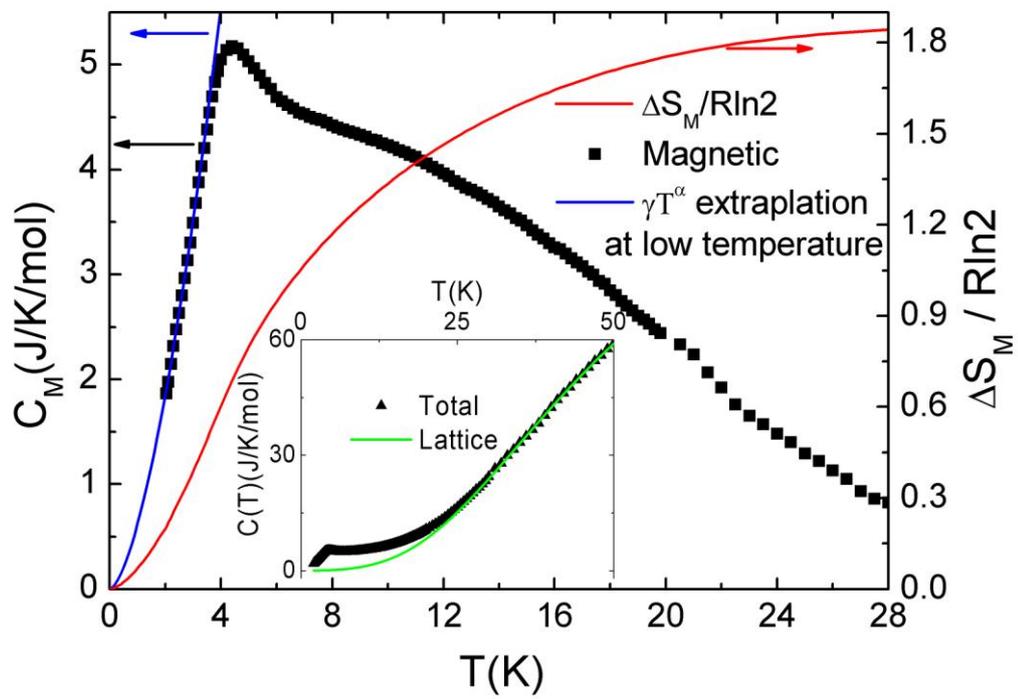

**Fig. 4.** Temperature dependence of magnetic specific heat and increasing magnetic entropy for CoCu$_3$(OH)$_6$Cl$_2$. The insert shows the total and lattice specific heat.

**Table 1.** Relative intensities of Bragg reflections at three incident photon energies, structure factors and non-resonant atomic scattering factors of Cu and Co

| h k l | Q (Å$^{-1}$) | M$^a$ | Observed relative intensity (%) | | | $\beta_t$$^b$ | $\beta_k$$^b$ | $\sum \beta f_R$$^b$ | $f_0^{Co}/f_0^{Cu}$$^c$ |
| --- | --- | --- | --- | --- | --- | --- | --- | --- | --- |
| | | | Co K-edge | 12.7keV | Cu K-edge | | | | |
| 1 0 1 | 1.150 | 6 | 80.2 | 91.4 | 77.1 | -3 | -3 | 1.97 | 17.63/19.79 |
| 0 0 3 | 1.338 | 2 | 21.4 | 17.3 | 5.0 | -3 | 9 | -19.16 | 16.88/18.95 |
| 0 1 2 | 1.386 | 6 | 10.2 | 2.8 | 0.8 | 3 | -3 | -24.51 | 16.70/18.75 |
| 1 1 0 | 1.837 | 6 | 3.1 | 5.6 | 11.5 | 3 | -3 | 39.97 | 15.19/17.07 |
| 0 2 1 | 2.167 | 6 | 39.2 | 31.9 | 13.9 | -3 | 9 | 9.15 | 14.27/16.04 |
| 1 1 3 | 2.272 | 12 | 100 | 100 | 100 | -3 | -3 | -45.51 | 14.01/15.74 |
| 2 0 2 | 2.301 | 6 | 11.3 | 21.3 | 5.2 | 3 | 9 | -74.57 | 13.94/15.66 |
| 0 0 6 | 2.675 | 2 | 11.7 | 15.2 | 7.5 | 3 | 9 | -4.81 | 13.09/14.71 |
| 0 2 4 | 2.771 | 6 | 88.7 | 92.9 | 73.0 | 3 | 9 | 35.77 | 12.90/14.49 |
| 2 1 1 | 2.841 | 12 | 1.7 | 7.4 | 1.7 | -3 | -3 | 39.08 | 12.76/14.33 |
| 2 0 5 | 3.077 | 6 | 20.5 | 14.5 | 6.4 | -3 | 9 | 21.34 | 12.32/13.83 |
| 1 0 7 | 3.296 | 6 | 15.3 | 18.1 | 14.6 | -3 | -3 | -31.87 | 11.95/13.40 |
| 3 0 3 | 3.451 | 12 | 37.2 | 39.7 | 37.0 | -3 | -3 | -17.98 | 11.70/13.12 |
| 2 2 0 | 3.673 | 6 | 49.7 | 47.3 | 39.3 | 3 | 9 | 59.83 | 11.37/12.74 |
| 0 2 7 | 3.773 | 6 | 8.9 | 6.5 | 1.9 | -3 | 9 | 1.49 | 11.23/12.58 |

$^a$ Number of equivalent crystal planes. $^b$ $\beta_t f_t$, $\beta_k f_k$ and $\sum \beta f_R$ denote structure factors of triangular, kagome and other sites (O and Cl). $^c$ Q-dependent non-resonant atomic scattering factors of Co/Cu[23].

**Appendix A**

**Table A.** Parameters of combined Rietveld refinements using theoretical delta-form factors at different photon energies. [a]

| R-3m model | | Combined/total | Co K-edge | 12.7keV | Cu K-edge |
|---|---|---|---|---|---|
| Delta-form Factors [b] | $f_{Co}'$ | - | -7.790 | 0.235 | -0.849 |
| | $f_{Co}''$ | - | 0.475 | 1.707 | 3.032 |
| | $f_{Cu}'$ | - | -1.686 | -0.014 | -7.854 |
| | $f_{Cu}''$ | - | 0.639 | 2.196 | 0.483 |
| | $f_{O}'$ | - | 0.053 | 0.022 | 0.041 |
| | $f_{O}''$ | - | 0.035 | 0.012 | 0.026 |
| | $f_{Cl}'$ | - | 0.373 | 0.233 | 0.335 |
| | $f_{Cl}''$ | - | 0.760 | 0.297 | 0.574 |
| formula ×3 | | $Co_{0.992(10)}Cu_{3.008(10)}(OH)_6Cl_2$ | \ | \ | \ |
| Lattice Parameters | a = b (Å) | 6.841182(175) | \ | \ | \ |
| | c (Å) | 14.092567(409) | \ | \ | \ |
| Cell volume (Å$^3$) | | 571.193(39) | \ | \ | \ |
| Co/Cu ×3 [-3M(100)] | Frac [c] | 0.800(10)/0.200(10) | \ | \ | \ |
| | 100*$U_{iso}$ [d] | 1.012(76) | \ | \ | \ |
| Cu/Co ×9 [2/M(100)] | Frac | 0.936(10)/0.064(10) | \ | \ | \ |
| | 100*$U_{iso}$ | 1.209(49) | \ | \ | \ |
| O ×18 [M(110)] | x = - y [e] | 0.203725(181) | \ | \ | \ |
| | z [e] | 0.061372(173) | \ | \ | \ |
| | 100*$U_{iso}$ | 0.844(95) | \ | \ | \ |
| Cl ×6 [3M(100)] | z | 0.194473(122) | \ | \ | \ |
| | 100*$U_{iso}$ | 1.354(64) | \ | \ | \ |
| Fitted [f] | wRp | 0.0497 | 0.0576 | 0.0664 | 0.0384 |
| | Rp | 0.0365 | 0.0426 | 0.0503 | 0.0293 |
| -Bknd [g] | wRp | 0.0691 | 0.0683 | 0.0900 | 0.0609 |
| | Rp | 0.0529 | 0.0564 | 0.0678 | 0.0428 |
| R (F$^2$) | | - | 0.0370 | 0.1841 | 0.0856 |
| Scan range (2θ$^o$) | | - | 10 - 120 | 5 - 120 | 10 - 120 |
| Nobs [h] | | 727 | 113 | 441 | 173 |

Reduced $\chi^2$ = 1.925 for 25 variables

[a] Considering the contribution of proton is insignificant, the positional parameters of proton are fixed in the refinements [6]. [b] The experimental x-ray incident energy of 8978 eV exactly locates at the K-edge of atomic Cu, according to the table given on the website "http://skuld.bmsc.washington.edu/scatter/AS_periodic.html". Considering the oxidation state

of Cu in $CoCu_3(OH)_6Cl_2$, the actual absorption edge will be lifted up by ~ 10 eV [30]. It means that the experimental incident X-ray energy is effectively lower than the actual absorption edge of $Cu^{2+}$ ions in $CoCu_3(OH)_6Cl_2$. Therefore, in our case we can safely use the imaginary part below the absorption edge of atomic Cu, i.e. 0.483. [c] Site occupation fraction. [d] Thermal motion parameter. [e] Fractional coordinates. [f] Refinement residuals covering the whole measured range. [g] Refinement residuals only with the contributions of Bragg reflections. [h] Number of observed reflections.